\documentclass{ws-procs10x7}

\begin{document}

\title{\hfill IFT-04/022\\
\hfill IPPP/04/60\\
\hfill DCPT/04/120\\[5mm]
TESTING CP VIOLATION AND UNIVERSAL EXTRA DIMENSIONS AT FUTURE COLLIDERS
} 

\author{JAN KALINOWSKI}

\address{ITP, Warsaw University, Ho\.za 69, 00681
  Warsaw, Poland \\
IPPP, University of Durham, Durham DH1 3LE, UK\\
E-mail: Jan.Kalinowski@fuw.edu.pl}

\twocolumn[\maketitle\abstract{ A future linear $e^\pm e^-$ collider with a
  clean environment, tunable collision energy, high luminosity, polarized
  incoming beams, and additional $e^-e^-$, $e\gamma$ and $\gamma\gamma$ modes,
  will offer precision tools to explore new physics. Here we summarize
  three papers submitted to the ICHEP04 conference in which polarized $e^+e^-$
  and $\gamma\gamma$ beams are exploited to search for CP violation, and
  universal extra dimensions (UED).}]

%
\section{Transverse beam polarization at a linear $e^+e^-$ 
collider and new physics$^{1,2,3}$} 
A 1 TeV linear $e^+e^-$ collider with high luminosity and 
polarized beams is now a distinct possibility.  If spin
rotators can produce transversely polarized beams (TPB), 
providing two more vectors, they would make it possible to observe
CP violation by observing a single final-state particle without measuring its
spin. Two specific processes 
have been considered:\\[1mm] 
\underline{ a) {$e^+e^-\to t\bar t$\cite{R1,R2}}}\label{subsec:prod}\\
The four-Fermi Lagrangian of beyond the SM contact $e\bar e t\bar t$
interactions (CI), after Fierz transformation, takes the form
\begin{eqnarray}\label{lag4f}
{\cal L}^{4F}
 =\Sigma_{ij}[\: S_{ij}(\bar{e}e)_i(\bar{t}t)_j+ 
 V_{ij}(\bar{e}\gamma_{\mu}e)_i(\bar{t}\gamma^{\mu}t)_j\nonumber \\
+ T_{ij}
 (\bar{e}{\sigma_{\mu\nu}}e)_i
(\bar{t}{\sigma^{\mu\nu}}t)_j\:]\mbox{\rm ~~~~~~~~~~~~~~}
\end{eqnarray}
where $(\bar{f} {\cal O}f)_i$=$ \bar{f}{\cal O} P_i f$, $S_{RR}$=$S^{*}_{LL}$,
$V_{ij}$=$V^{*}_{ij}$, $T_{RR}$=$T^{*}_{LL}$ and non-diagonal $S$ and $T$
vanish.

\begin{figure*}
\epsfxsize30pc
\figurebox{16pc}{32pc}{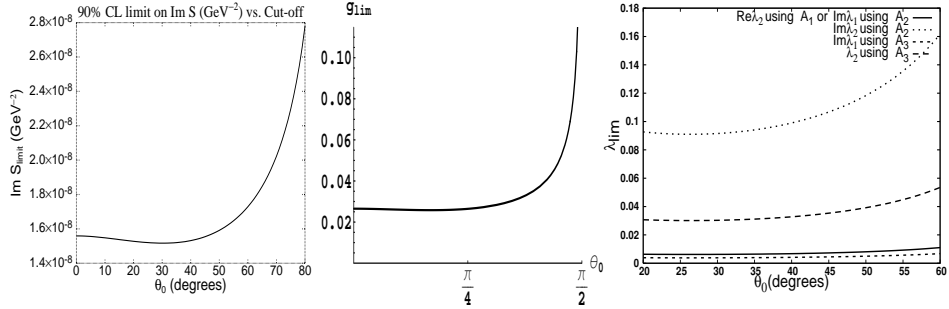}
\caption{The 90\% CL limits on: Im $S$ (left), leptoquark (center), and
  $\gamma\gamma Z$ and $\gamma ZZ$ couplings (right), 
as functions of the cut-off angle
  $\theta_0$.  [$\sqrt{s}$=500 GeV, $\int Ldt$=500 fb$^{-1}$]} \label{figR}
\end{figure*}

With TPB the chirality-violating $S$ and $T$ terms interfere with the SM
contributions and can be studied, in contrast to the cases of 
no or longitudinal beam
polarization, where they appear only at second order.  Assuming $P_{e^-}$
along the $x$-axis and anti-parallel $P_{e^+}$, the CP-odd azimuthal
asymmetry, defined as
\begin{equation}
A_1(\theta_0)=\frac{N(\sin\phi>0)-N(\sin\phi<0)}{N(\sin\phi>0)+N(\sin\phi<0)},
\label{eq:asym}
\end{equation}
is sensitive to 
\begin{equation}\label{S}
{\rm Im} S\equiv {\rm Im}(S_{RR} + {
\frac{2c_A^t c_V^e}{ c_V^t
    c_A^e} T_{RR}} )
\end{equation}
Here $N$ is the number of events with a given azimuthal angle $\phi$ and polar
angle of the top quark within the $\theta_0$ cut as
$\theta_0<\theta<\pi-\theta_0$.  Fig.\ref{figR} (left) shows the 90\% C.L.
limit on ${\rm Im}S$ for an integrated luminosity of $\int Ldt
=500$ fb$^{-1}$ and
$\sqrt{s}=500$ GeV and 100\% TPB. The limit of $1.6\cdot 10^{-8}$ TeV$^{-2}$
translates to a scale $\Lambda$ of CI of order 8 TeV (for $\sqrt{s}$ of 800
GeV the sensitivity increases to $\sim 9.5$ TeV).  For realistic TPB of
$P_{e^-}=0.8$ and $P_{e^+}=0.6$, the limit on $\Lambda$ goes down to about 6.7
TeV.

A specific example of chirality-violating couplings at tree level has been
considered in\cite{R2} with an $SU(2)_L$ doublet of scalar leptoquarks coupled
only to first-generation leptons and third-generation quarks. The azimuthal
asymmetry $A_1$ turns to be proportional to ${\rm Im}(g_Rg_L^*)$ while an
asymmetry $A_2$, defined as in Eq.(\ref{eq:asym}) but with $\phi \to
\phi+\pi/2$ and with parallel $P_{e^+}$ and $P_{e^-}$, is proportional to
${\rm Re}(g_Rg_L^*)$; $g_i$ is the leptoquark coupling to the chiral current
$(\bar t e)_i$.  Fig.\ref{figR} (center) shows the 90\% C.L. limits on ${\rm
  Re}(g_Rg_L^*)$ and ${\rm Im}(g_Rg_L^*)$ for leptoquark mass of 1 TeV with
realistic TPB.  The limit on ${\rm Re}(g_Rg_L^*)$ is competitive with $|{\rm
  Re}(g_Rg_L^*)|<0.1$ derived from $g_e$--2, while the limit $|{\rm
  Im}(g_Rg_L^*)|<10^{-6}$ from the electron EDM is much
stronger.\\[2mm]
\underline{b) {$e^+e^-\to \gamma Z$\cite{R3}}}\\ 
Here the final-state particles are both self-conjugate. With TPB, a
T-odd, CP-even azimuthal asymmetry can be combined with the T-even, 
CP-odd forward-backward (FB) asymmetry to give an asymmetry which is both CP-
and T-odd. A CP-vio\-lating contribution can arise if anomalous CP-violating
$\gamma\gamma Z$ and $\gamma ZZ$ couplings are present.

The effective CP-violating Lagrangian for $\gamma\gamma Z$ and $\gamma ZZ$
interactions, up to dimension 6 terms, can be written as
\begin{eqnarray}
{\cal L} =  \displaystyle
   \frac{e\, \lambda_1}{ 2 m_Z^2} F_{\mu\nu}
    \left( \partial^\mu Z^\lambda \partial_\lambda Z^\nu
          - \partial^\nu Z^\lambda \partial_\lambda Z^\mu
      \right) \nonumber
       \\
 \displaystyle
      +\frac{e\, \lambda_2}{16 c_W s_W m_Z^2}
       F_{\mu\nu}F^{\nu \lambda}
       \left(\partial^\mu Z_\lambda + \partial_\lambda Z^\mu   \right)
 {\rm ~~}
      \label{lagrangian}
\end{eqnarray}
To isolate appropriate anomalous couplings three different CP-odd asymmetries,
which combine a FB asymmetry with an appropriate asymmetry in $\phi$, have
been identified in\cite{R3}.  The derived 90\% C.L. limits on real and
imaginary parts of the $\lambda_1$ and $\lambda_2$ couplings, plotted as
functions of the cut-off $\theta_0$, are shown in the right panel of
Fig.\ref{figR}.

%
\section{Resonant H/A  mixing in the CP-noninvariant SUSY$^4$}
The tree level Higgs potential of the MSSM is CP-conserving implying two $h,H$
of the three neutral states to be CP-even, while the third $A$ is CP-odd. With
non-vanishing CP phases in the soft SUSY-breaking terms, however, radiative
corrections induce the three neutral bosons to mix forming a triplet
($H_1,H_2,H_3$) with even and odd components in the wave-functions under CP
transformations.

As expected from quantum mechanical rules, the mixing can become very
large if the states are nearly mass-degenerate. This situation is naturally
realized in the decoupling limit in which two of the neutral states, $H$ and
$A$, are heavy. The lightest Higgs $H_1$ then becomes the CP-even SM-like
Higgs, and does not mix with the $H/A$ system.  In this limit the off-diagonal
mixing term of the 2$\times$2 mass matrix $M^2$ in the $H,A$ basis reads
\begin{equation}
M^2_{HA} =v^2[ {\textstyle \frac{1}{2}}\lambda_5^I c_{2\beta}
                    -{\textstyle \frac{1}{2}} 
(\lambda_6^I-\lambda_7^I) s_{2\beta}]
\end{equation}
where $\lambda^I_i$ are imaginary parts of loop-induced quartic Higgs
couplings\cite{C1}.  

For small mass differences, the mixing is strongly affected by the widths of
the states and the complex, symmetric Weisskopf--Wigner mass matrix ${\cal
  M}_c^2=M^2-iM\Gamma$ must be considered in total, not
only the real part. 
Recently a coupled-channel method has been employed\cite{ELP} for the Higgs
formation and decay processes at the LHC. 

In Ref.\cite{C1} an alternative approach has been followed where the full mass
matrix ${\cal M}_c^2$ is diagonalized
\begin{equation}
{\cal M}_{\rm diag}^2=C {\cal M}_c^2 C^{-1}
\end{equation}
For the $H/A$ system, the {\it complex} 2$\times$2 rotation matrix is
expressed in terms of a {\it complex} mixing angle $\theta$, which is given by
\begin{equation}
X=\frac{1}{2}\tan2\theta=\frac{{M}^2_{HA}-iM_{HA}\Gamma_{HA}}
{{\cal M}^2_{HH}-{\cal
    M}^2_{AA}} 
\end{equation}
where ${\Gamma}_{HA}$ (${\cal M}^2_{HH},{\cal M}^2_{AA}$) is the off-diagonal
(diagonal) entry of the decay (complex mass ${\cal M}_c^2$) matrix.
\begin{figure}
\epsfxsize200pt
\figurebox{1pt}{2pt}{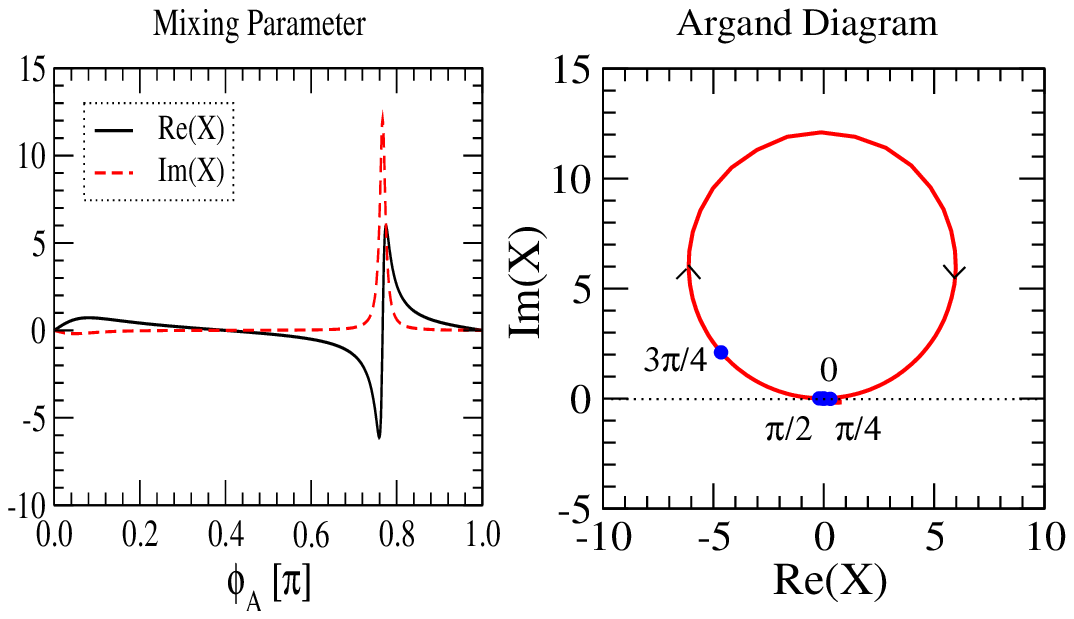}
\caption{The $\phi_A$ dependence of
     the mixing parameter $X$ (left) and the Argand diagram (right) 
in a SUSY model with the
     CP-violating phase $\phi_A$.}
\label{fig:arg}
\end{figure}

In Fig.\ref{fig:arg} the complex $H/A$ mixing in the MSSM is shown for a
typical set of parameters: $M_S$=0.5 TeV, $|A_t|$=1 TeV, $\mu$=1 TeV,
$\tan\beta$=5, while varying the phase $\phi_A$ of the trilinear parameter
$A_t$.

A future photon collider would be an ideal tool to study resonant CP-violation
in the Higgs sector. Two promising signatures have been
considered in Ref.\cite{C1}:\\[1mm]
\underline{a) {\it $\gamma\gamma \to H_i$ formation with polarized beams}}\\
For linearly polarized photons, the CP-even component of the $H_i$
wave-functions is projected out if the polarization vectors are parallel, and
the CP-odd component if they are perpendicular.  This can be observed in the
CP-even asymmetry ${\cal A}_{lin}$, 
\begin{eqnarray}
{\cal A}_{lin}
  = \frac{\sigma_\parallel- \sigma_\perp}{\sigma_\parallel+ \sigma_\perp},
  \quad 
{\cal A}_{hel}
  =\frac{\sigma_{++} -\sigma_{--}}{\sigma_{++}+ \sigma_{--}},{\rm ~~}
\label{ggasym}
\end{eqnarray}
since $|{\cal A}_{lin}|$$<$1 requires both scalar and pseudoscalar
$\gamma\gamma H_i$ non-zero couplings.  Moreover, CP-violation due to $H/A$
mixing can directly be probed via the CP-odd asymmetry ${\cal A}_{hel}$
constructed with circular photon polarization, as
defined in Eq.(\ref{ggasym}).\\[1mm]
\begin{figure}
\epsfxsize200pt
\figurebox{1pt}{2pt}{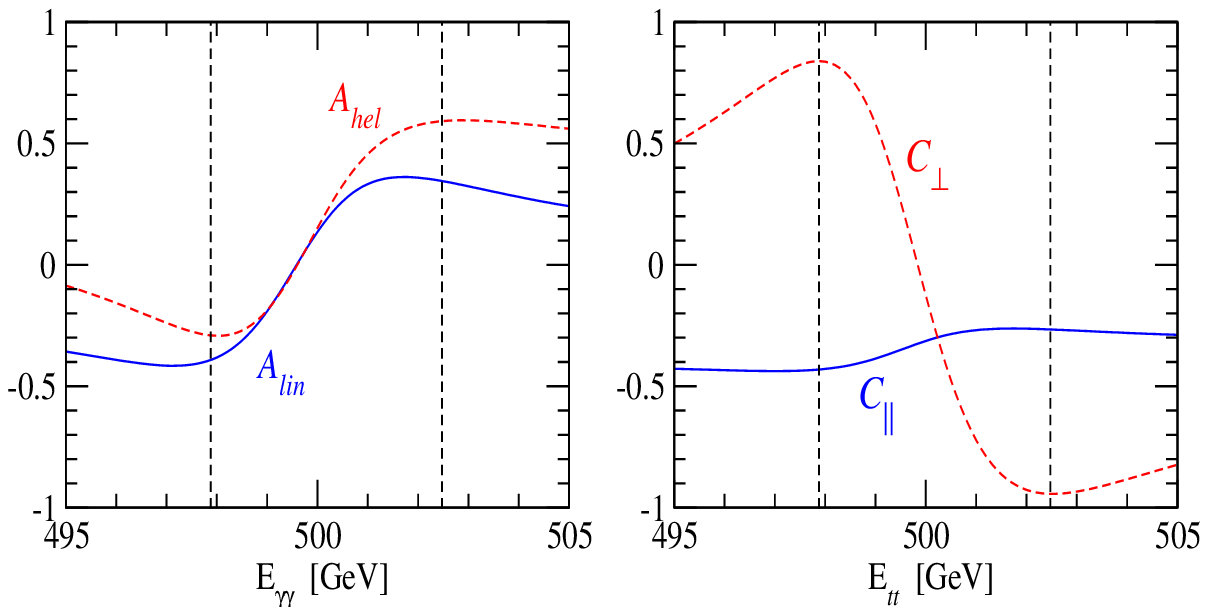}
\caption{Left: The
         $E_{\gamma\gamma}$ dependence of  
         ${\cal A}_{lin,hel}$  in the production process
         $\gamma\gamma \rightarrow H_i$. Right:  The 
         $E_{t\bar{t}}$  dependence  of 
         ${\cal C}_{\parallel,\perp}$ in the
         production--decay chain $\gamma\gamma \rightarrow H_i\rightarrow
         t\bar{t}$. In both figures $\phi_A=3\pi/4$. The vertical
lines represent the two mass eigenstates.}
\label{fig:asym}
\end{figure}
\underline{b) {\it polarization of top quarks in $H_i$ decays}}\\ 
%
CP-induced correlations between the transverse $t$ and $\bar{t}$
polarization vectors $s_{\bot},\bar{s}_{\bot}$ in the decay process
$H_i\rightarrow t\bar{t}$,
\begin{eqnarray}
{\cal C}_\parallel = \left\langle s_\perp \cdot \bar{s}_\perp
                 \right\rangle, \quad
{\cal C}_\perp = \left\langle \hat{p}_t\cdot (s_\perp\times\bar{s}_\perp)
                 \right\rangle, \quad
\end{eqnarray}
lead to a non--trivial CP-even and a CP-odd azimuthal correlation,
respectively, between the decay planes of $t\to bW^+$ and $\bar{t}\to
\bar{b}W^-$.

The left panel of Fig.\ref{fig:asym} shows the asymmetries ${\cal A}_{lin}$
(blue solid line) and ${\cal A}_{hel}$ (red dashed line) in the $\gamma\gamma
$ collider as the $\gamma\gamma$ energy is scanned from below $M_{H_3}$ to
above $M_{H_2}$. The right panel shows the $E_{t\bar{t}}$ dependence of the
correlators ${\cal C}_\parallel$ (blue solid line) and ${\cal C}_\perp$ (red
dashed line). Both figures are for $\phi_A=3\pi/4$, a phase value close to
resonant CP-mixing.  Detailed experimental simulations would be needed to
estimate the accuracy with which they can be measured. However, the large
magnitude and the rapid, significant variation of the CP-even and CP-odd
asymmetries through the resonance region would be a very interesting effect to
observe in any case.

\def\ltap{\raisebox{-.4ex}{\rlap{$\sim$}} \raisebox{.4ex}{$<$}} 
\def\gtap{\raisebox{-.4ex}{\rlap{$\sim$}} \raisebox{.4ex}{$>$}} 
\section{Testing the UED at a linear $e^+e^-$ collider$^6$} 
In the simplest UED scenario considered in\cite{K1} one extra dimension is
accessed by all SM particles and low-energy data constrain the
compactification radius $R^{-1}~\gtap$ a few hundred GeV. The extra dimension
$y$ is compactified on a $S^1/Z_2$ orbifold, rendering all matter and gauge
fields, viewed from 4d, dependent on $y$ either as $\cos(ny/R)$
or $\sin(ny/R)$, where $n$ is the corresponding KK number. The mass of the
$n$th KK state is given by $M_n^2 = M_0^2 + n^2/R^2$, where $M_0$ is the zero
mode mass of that field. A remnant $Z_2$ symmetry
dictates that the KK parity defined as $(-1)^n$ is conserved, implying (i) the
lightest Kaluza-Klein particle (LKP) is stable, and (ii) a single KK state
cannot be produced --  a reminiscent of supersymmetry with
conserved R-parity.

A tree-level degeneracy of the KK modes of light SM particles is lifted by
radiative corrections: finite bulk corrections $\Delta M_n^2 \propto
\beta/16\pi^4 R^2$, and logarithmically divergent orbifold corrections $\Delta
M_n \sim M_n (\beta/16\pi^2) \ln(\Lambda^2/\mu^2)$; $\beta$ represents the
$\beta$-functions of the gauge and matter KK fields in the loop, and $\Lambda$
is the UV cut-off scale.
\begin{figure}
\epsfxsize13pc
\figurebox{6pc}{2pc}{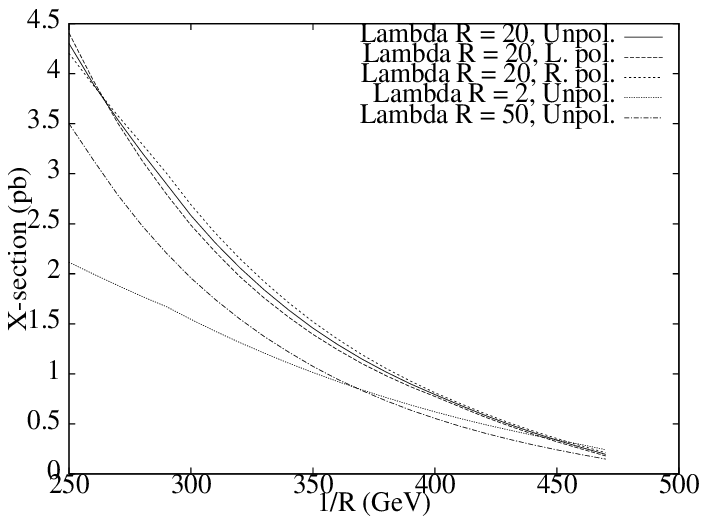}
\caption{$1/R$ dependence of $\sigma(e^+e^-\to 
e^+e^- +  E_{\rm miss}$ for unpolarized 
beams with $\Lambda R = 2,20$ and 50. Polarization 
effects on the $e^-$ beam are shown for $\Lambda R=20$. 
 \label{fig:K}}
\end{figure}

The first KK modes of electrons, $e^\pm_1$, should be copiously produced in
$e^+e^- \to e_1^+ e_1^-$ at a future linear collider.  The process proceeds
through the $s$-channel $\gamma$ and $Z$, and the $t$-channel $\gamma_1$ and
$Z_1$ exchanges.  The splitting between $e_1$ and $\gamma_1$ turns to be
sufficient for the decay $e^-_1\to e^-+\gamma_1$ with $\gamma_1$ likely to be
the LKP and escaping detection.  So the final state is $e^+e^-$ and 2
$\gamma_1$'s carrying away missing energy.  The $W_1^\pm$ pair production with
$W_1\to e +\nu_{e 1}$, $e_{1}+\nu_e$ is numerically insignificant.  The main
SM background from $W^+W^-$ 
pair production 
can be eliminated by requiring
that the final electrons are sufficiently soft.

Fig.\ref{fig:K} shows the cross section for $e^+e^-\to e^+e^-+E_{\rm miss}$
with $e^\pm$ energies between 0.5 and 20 GeV and polar angle away from the
beam pipe by $15^\circ$.  Events coming from excited $W_1$ have been
neglected.  Not much is gained from the beam polarization.  The cross section
enhances as $\Lambda R$ is increased from 2 to 20 due to the change in the
$n=1$ Weinberg angle. Clearly the signal events are quite prominent, and the
SM background after cuts is under control.

\section{Conclusions}
The examples discussed above add an additional weight to the physics case of a
linear collider, strengthening its ability to explore physics beyond the
standard model.

\vskip 3mm 
\noindent {\bf Acknowledgments}\\[1mm]
Thanks go to S.Y.Choi, A.Kundu, Y.Liao, 
S.Rindani, J.Stirling and P.Zerwas for useful
comments. 
Work supported by the Polish Committee for Scientific Research (KBN) 
grant 2 P03B 040 24 (2003-2005).

\end{document}